\documentclass[pra,showpacs,showkeys,amsfonts,12pt]{revtex4}
\usepackage{graphicx}
\RequirePackage{times}
\RequirePackage{mathptm}
\RequirePackage{textcomp}
\RequirePackage[german]{babel}
\RequirePackage[isolatin]{inputenc}
\begin{document}
\title{Peer review in context\footnote{
Presented at the conference
{\em ODOK '03}, in Salzburg, Austria, September 23-26, 2003
(see {\tt http://voeb.uibk.ac.at/odok2003/svozil.pdf} for a German version),
and at the
{\em INST} conference {\em The Unifying Aspects of Cultures}
Vienna, November 7-9, 2003.
The opinions expressed in this article
should be understood as a subjective evaluation of and contribution to an
ongoing discussion. They
do not reflect the official position of the Vienna University of Technology. }}
\author{Karl Svozil}
\email{svozil@tuwien.ac.at}
\homepage{http://tph.tuwien.ac.at/~svozil}
\affiliation{Institut f\"ur Theoretische Physik, University of Technology Vienna,
Wiedner Hauptstra\ss e 8-10/136, A-1040 Vienna, Austria}

\begin{abstract}
Scientific publishing is in a transition between the
old paper-bound, static forms and the new electronic media with its
interactive, dynamic possibilities.  This takes place in the context of
imploding library budgets and exploding magazine costs.  The scientists
as authors, reviewers and editors of scientific journals are exposed to an
increased pressure by the their administrations and the public
towards quantification, objectification and certification of scientific
achievements.  The ``publication roulette'' resulting from
low-quality editorial procedures often amounts to malign
censorship, which not only is experienced as a frustration by the authors,
but is also delaying and hampering the progress of science.
It also leads to a
waste of funds under the cover of pseudo-objectivity and pseudo-legitimacy of financial decisions.
Different solutions are outlined
and discussed.  As concerns scientific publishing, an e-print service
should be established, which, in continuation of existing e-servers such
as {\it arxiv.org}, is operated either directly by the United Nations Educational, Scientific and Cultural
Organization, or by an international consortium.  In order to become generally accepted by
the scientists, certification criteria must be provided, which would
make it possible to successfully pursue a scientific career besides the
traditional peer reviewed print publications.
\end{abstract}

\maketitle
%\newpage

%\tableofcontents

\section{Peer Review}

To a noninvolved observer, peer review can be explained as a kind
procedural pattern or ritual, in which a decision over the publication
of scientific reports (and/or over the funding of some research
project) is reached.  At the beginning, an unsolicited article is
submitted by the author about some research results.  The article is
sent from the editor to unpaid reviewers, called {\it peers}.  These
reviewers provide reports and recommendations which are sent back to the
editor.  The editor anonymizes the reports and sends them to the
authors.  The article is revised by the author and re-submitted.  This
procedure can repeat itself.  Finally, the editor decides whether or not
the article is worth publishing or is rejected.  Rejections rates vary
strongly, depending on the field covered, from 10 \% to 95 \%.
And despite
the critical evaluation of the situation, most protagonists attempt to
do a decent job under the given circumstances.

\subsection{Why peer review?}

Peer review has at least three main goals:
(i) quality certification of scientific publications, (ii)
career planning of the new scientific generation by comprehensible,
``objective,'' quantitative criteria, as well as (iii) the evaluation of
 research projects requesting funding .

The importance of peer review for scientific careers is enormous:  a
publication which does not appear in a journal whose contributions are
subjected to peer review, is mostly considered ``worth nothing'' in
terms of career planning;  and without peer review there is no certified
progress in science; at least this is what is emphasized over and over
again.  Therefore, it is mandatory for the novices as well as for the
established researchers requesting positions, status, influence and
resources, to expose themselves to this verdict.  And although most
authors express their frustration with this kind of
censorship behind closed doors, public criticism is considered inappropriate; except if
one is willing to bear the consequences, such as being denoted a
``whiner.''

Peer review is seen primarily as assistance to the author for improving
articles.  It avoids the publication of uninteresting, plagiaristic,
faulty, erroneous and fake results.
Each reader should form an own judgment whether or not these advantages,
should they be achieved, counterbalance the disadvantages of the
scientific censorship.  These
issues deserve public concern.  After all, not to a small part tax money
and the pursuit of scientific progress is at stake.

\subsection{Peer review in the historical perspective}

The history of peer review  needs still to be written;
amazingly few details have been documented.  After 1650, the first
magazines of scientific societies developed, whose members understood
themselves as ``peers.''  Examples are the {\it Institutes de France} or
the {\it Royal Society of London,} publishing magazines like the {\it
Journal des Savants} or the {\it Philosophical Transactions,} which
already used the review process among the ``peers'' for editorial
purposes.  In which form this happened does not seem to have been sufficiently
examined yet.

The fact that already in the early stages the system needed improvements
and adjustments is documented by a quotation of Babbage around 1830
\cite{babbage-1830}:  \begin{quote} {\em $\ldots $ it would be a
material improvement on the present mode, if each paper were referred to
a separate Committee, who should have sufficient time given them to
examine it carefully, who should be empowered to communicate on any
doubtful parts with the author; and who should report, not only their
opinion, but the grounds on which that opinion is formed, for the
ultimate decision of the Council.} \end{quote} No reference is given to
the necessity of and the reasons for anonymity; as well as to the costs of
this procedure.

Often one hears the astonishment in physical circles over the
willingness of one of the most outstanding physics journals of its time,
i.e., the {\it Annalen der Physik} and his editor R\"ontgen, to publish
the groundbreaking ideas of a hitherto unknown official in the Swiss
patent office called Einstein.  Would such a thing still be conceivable
today?  Einstein's attitude towards peer review, as he experienced it in
the USA, can probably be best characterized by an anecdote mentioned in
Pais' biography of Einstein  \cite[pp. 494-495]{pais}.  In 1937, Einstein
had submitted an article to the {\it Physical Reviews}
and got back a lengthy review.  His immediate reaction appears
unexpected to contemporary scientists:  \begin{quote} {\em Einstein was
enraged and wrote to the editor that he objected to his paper being
shown to colleagues prior to publication.  The editor courteously
replied that refereeing was a procedure generally applied to all papers
submitted to his journal, adding that he regretted Einstein may not have
been aware of this custom.  Einstein sent the paper to the Journal of
the Franklin Institute and, apart from one brief note of rebuttal, never
published in the Physical Review again.} \end{quote}

A further anecdote is about the joke which the later Nobel laureate Bethe
made by ridiculing Eddington's inclinations for numerology in deriving
the fine structure constant from the absolute zero point temperature;
this article got published an  in
{\it Die Naturwissenschaften} in the year 1931
\cite{beck-bethe-riezler}.

The so-called ``Sokal affair'' \cite{sokal-aff}, in which the New York
physicist Sokal ridiculed the publication efforts in the Social
Sciences, is already a legend.  (Marketing-wise, i.e., in terms of
self-promotion, this conscious fraud of Sokal is a great achievement; it
moved up the then widely unknown Sokal into the center of the scientific
and even general attention; Sokal was discussed, considered and invited
everywhere.)  The arrogance of this gentleman became clear, when physics
afterwards suffered from affairs of her own; beginning with the frauds
of Sch\"on and his co-authors \cite{schoen-2002}, to the asseverations
of the Bogdanov brothers to have manufactured their thesis and
articles, which later were certified and published in venerable peer
reviewed journals, in earnestness and good scientific conduct
\cite{bogdanov-affair}.  To well camouflaged scientific fraud and
charlatans probably the same applies as to perfect crimes:  they remain
mostly hidden.  At the moment, there are no estimates of
the estimated number of unknown cases
of such occurrences.

It would be probably a worthwhile task to examine the angloamerican
influence on the scientific publication regime in the time after the
World War II.  Here one may express the assumption that in Central
Europe everything changed dramatically; and that the American peer
review model became generally accepted; to a degree, which
makes it almost impossible to reconstruct the traditions and practices
of the time before the World Wars.

Nowadays completely different signals are sent from the USA, once again
showing the strength and innovative potential of this great nation:
motivated by the necessity of rapid dissemination of research results in
high-energy physics, an area rapidly developing in the nineties, an
electronic system of ``preprint'' or ``reprint servers'' developed
``bottom-up,'' which today has become a {\it de facto} standard, and the
main distribution channel of scientific literature:  we shall deal with
{\it arxiv.org} later on in greater detail.

\section{Peer review ``from the inside''}

Scientists experience peer review in three different functions:  (i) as authors,
(ii) as reviewers, and (iii) as editors.
In what follows, these functions will be dealt with
briefly.  The reader is also referred to the publications of Fr\"ohlich
\cite{2002-froehlich,2003-froehlich}, which offer a wealth of thoughts,
details, investigations and much background information on this topic
(see also Ref.  \cite{sala-2002} and the following articles of the
magazine {\it Cortex}).

\subsection{Authorship and ``Publication roulette''}

As already described, an author, who wants to publish the results of its
scientific work, writes an article and sends it, usually
electronically, to the editor of a scientific journal.  After a more or
less long waiting period, the author receives a reaction, which depends
on the recommendation of the reviews.  To a certain extent, the author experiences
a ``supply sided market situation:''  there are always
sufficiently many unsolicited articles between which a journal editor seems to be able
to select.

The referee reports are not always drawn up in a generous, respectful
style of benevolent criticism.  Indeed, often sarcastic, hurting and not
very sober, unobjective, even humiliating remarks of the ``peers'' are
passed on one-to-one from the editor to the author.  Unfortunately, this
habit, caused by unqualified, weak editors, which do not want to be bothered with
the quality of the review, but are just interested in evaluations, no
matter what, contributes much to a degeneration of manners in the
scientific community.  In the appendix, some anonymized anecdotes give a
sample of what could await an author of scientific articles trying to
publish them.

In a large-scale study \cite{1981-bradley} over 600 authors were asked  about
their experiences with peer review.  The results were devastating:  the authors
emphasized their frustration over peculiar reports, which criticized
unimportant details without dealing with the main results; they
emphasized the incompetence of the peers, who treated the authors
arrogantly.  Many authors suspected also that from their experience many
reports had not been written in order to improve the quality of the
articles but to impress the editors.

In the long run, sooner or later, almost every article succeeds to get
published by some peer reviewed journal.  One rumors about cascade-type
publication tactics, which begin with the submission to the most
prestigious journals and, in the case of refusal, continues with
less respected and less known journals  until final publication.

The time delay caused by the peer review procedure, also for manuscripts
which get immediately accepted, amounts from three months to several
years in extreme cases; on the average, the delay is about half until
three quarters of a year; at least in physics.  (Other fields might exhibit
still longer latencies.)  These not inconsiderable delays,
particularly in fast moving, active research areas, contributed
to the  development
of preprint servers, which by now have taken over almost an exclusive communication role
in scientific publishing.  In these fields, the additional,
``post-''publication in peer reviewed journals almost
exclusively for career planning is used.

From a szientometrical perspective, and for many fellow scientists and
administrations, the market value of the author is derived from the
market value of the journals in which this author publishes (as well as
from the outside funding attracted).  This procedure, which is often
related to indicators such as for instance the {\it Science Citation
Index (r)} [{\it SCI (r)}], which is copyright protected, owned and
operated by the private firm {\it Thomson/ISI}, and derivatives such as
the {\it impact factor}, results in problematic consequences and may
even lead to grotesque developments.  In general, for all kinds of quantitative
indicators, concerns are not completely unfounded that they
may lead to an inefficient form of scientific practice by adopting
marketing strategies to cope with the quantitative measures rather than
to concentrate on the quality of work:  Quantity instead of quality!

Nevertheless, and despite of all that, often serious suggestions,
comments and criticisms are conveyed by peer review, making the
manuscript better and preventing mistakes.  And some reports contain
so valuable suggestions that they would even justify co-authorship of
the anonymous peer.  The question remains whether the advantages
outweigh the disadvantages.

\subsection{Reviewer:  no time, no money}

In a large-scale study \cite{1981-cole}, 150 research projects of
physics, chemistry and economic science were re-examined by the {\it
National Science Foundation.} The results were devastating.
This study showed how strongly the acceptance or refusal of
a research project depends on the choice of the particular reviewer
evaluating that proposal:  \begin{quote} {\em An experiment in which 150
proposals submitted to the National Science Foundation were evaluated
independently by a new set of reviewers indicates that getting a
research grant depends to a significant extend on chance.''} They
proceed by stating that, {\em ``the degree of disagreement within the
population of eligible reviewers is such that whether or not a proposal
is funded depends in a large proportion of cases upon which reviewers
happen to be selected for it.} \end{quote}

Well into this picture fits a study, in which articles were re-submitted
after one and a half to three years to the same journals in which they were
already published \cite{1982-petersceci}.  Another issue is the tendency
of some reviewers to delay or even impede the publication of certain
competition articles for egotistic self-interests.

Still another problem is the bias, with which scientists judge their area in
comparison to others.  The Swiss {\it Wissenschaftsrat}, an advisor
committee of the government, attempted to gather the opinions of well
established scientists about prospective future research fields.  After
some years these recommendations were compared with the actual
developments.  Many recommendations were misleading.  Heinrich Urprung,
for many years the president of the {\it ETH Zurich} and Swiss secretary
of state for science and research, has expressed the findings as follows
\cite{ursprung-1990} (cf. also \cite{swizz-science}):

\begin{quote} {\em At the beginning of the seventies our science advisors
undertook a monumental effort to anticipate promising scientific
research areas of the future.  The strategy of this search consisted of
asking hundreds of professors about their opinion.  The result was an
impressive document with reference to urgently necessary upgrades of
those areas, which were already established at our universities.  In
this sense, the expenditure was worthwhile itself.  Meanwhile, at the
time, nearly nobody referred to the necessity for additional research in
semiconductor technology, and nearly nobody stressed the necessity for
increased efforts in the area of the energy research.  Either the
appropriate experts had not been reviewed, or the importance of the
development of their own areas escaped them.  Stated more generally,
the lack of such planning results from the fact that, by definition, gaps
do not have proponents.  In addition to that, professors, as many other
mammals and  most socially organized organisms, are characterized by
a pronounced territorial thinking.  } \end{quote}
In general,
speculations that peer review discriminates against innovative, not well
established ideas, and favors the advancement of extensions of contemporary knowledge,
do not appear
completely unfounded.  The latter would actually be nothing despicable,
but the first is problematic.

With all the respect, distrust and contempt,  brought forth for and against  peer
review, two further important factors should not be ignored, which
are essential to an understanding of the situation:  time and money.
Because on the one hand the pressure on single scientists from
committees and the administration gets bigger and bigger to submit as
many articles as possible.  On the other hand, as reviewers they are
expected to prepare their review assessments anonymous, by unpaid and
unnoticed.  So, there is a simple rule here:  a single article earns
more official bonuses than numerous referee reports.  The reviewing
efforts do not pay off; publishing houses and funding agencies assume
the (cheap) position that this work is an integral part of scientific
duty and paid off already be the scientist's salary.  This lack of
acknowledgment appears to be more absurd in the perspective of the rising
yields of scientific publishing houses (see below).

In particular, authors who are materially and organizational
insufficiently secured; which work with temporally limited contracts and
in various other dependencies, find themselves in a treadmill:  as
authors, they have to kindly accommodate their reviewers in order to get
their articles published; yet as reviewers they take pride to criticize
manuscripts and research proposals, without being able to reflect and
comprehend them sufficiently.  This observation is related to the finding that
younger reviewers recommend more
rejections than older peers \cite{1994-nylenna}.  Further studies
reflect on the discrimination of women in the reviewing process
\cite{wenneras-wold-97}, as well as on the dependence on the seniority
and status of the author (see also the appendix).

Maybe one way out of this malady would be the issuance of ``peer review
certificates'' by the publishing houses, which could be redeemed and
credited for career purposes by the reviewers in a sort of coupon
system.  This would also make it easier to analyze the distribution of
reviewers over the entire population of peers, which is another issue
not properly addresses in the current literature.  Maybe very few peers
do most of the reviewing, thereby maintaining a huge influence over what
appears to become scientific literature?   So far, no investigation
exists which tests this hypothesis.

\subsection{Editor}

If one considers the findings quoted above, according to which the
choice of the reviewers is crucial for the fate of an article or
research proposal, then the editor's role is central and influential.
It is amazing how little concern is given to the choice of the editors;
not only by the community of peers, but also by the political
institutions which provide funding.  An editor may ruin a journal or
fund, or may make it prosperous.  For instance, not much is known how
exactly the Austrian or European funding agencies select their executive
editors; and one is tempted to suspect that they are nominated to fit
the interests of important groups within the scientific communities.
Whether such procedures are optimal or even beneficial for the progress
of science is questionable; in particular if one recalls Ursprung's
verdict cited above.

Editors often have to ``chase down''
reviewers to write reports:  the reviewers must be kindly asked,
reminded and admonished repeatedly, until they deliver.  There are no
leverages despite moral obligations.

Complementary to the author, the editor experiences the
market situation ``demand sided;'' at least to a certain extent:  among
the submitted and received articles there are few very high-quality
ones.
As regards soliciting reports, there are often not too many
reviewers willing to seriously get involved with a manuscript.

Just as the author's and reviewer's role, the editor's role is
characterized by financial austerity.  However, this is to a certain
extend compensated by the editorial status, as well as by the influence
(in the ``benign'' sense;) exerted.

\section{Financial context}

The financial  conditions of scientific publishing are
characterized by imploding library budgets accompanied by exploding
journal costs.  The {\it Create} web page of the {\it Association of
Research Libraries} grants an eloquent insight into this situation.  A
case study describes the precarious situation in Australia with the
following numbers:  In the year 1993 had 38 university libraries in
Australia subscribed to altogether 200.666 science journals.  Until
1998, this number decreased to 112,974 subscriptions, a relative
decrease of 43.7 \%.  During this period, the cost of an average journal
increased from Australian \$ 287 to Australian \$ 485, a jump of 70 \%.

In another confrontation, the consumer price index rose in the period
from 1986 to 1998 by around 49 \%, while the average journal cost rose
around 175 \%, more than three times the increase of the consumer price
index.  In 1999, the American scientific libraries bought of 26 \% fewer
books than 1986.  During this time, world book production increased
by about 50 \%.

Also in Austria the situation appears precarious.  As an
example, consider the following numbers, which were recently sent to an Austrian
university institute, characterizing the situation of a typical Austrian
library budget in a snapshot:  ``sum planned (bound) expenditures:
259,345 EUR; literature budget assigned by the rector:  249,340 EUR;
Therefore, budget available for literature acquisitions:  -10,005 EUR.  (The
puzzled reader may ask if a negative library budget results in fewer
books.)

But the problems of one party often translates into the benefits of another:
The net profits of the commercial publishing houses marketing those
journals, have reach all all time high, and are still increasing; the profit
margins being higher than for fiction books.  For example, according to
data of the {\it Reed Elsevier Annual Report} the science publishing
division of {\it Reed Elsevier} has exhibited a profit margin of 35
\%-42 \% in the years 1995-1999.
One is tempted to view the science
libraries, which at least in Europe are to a large extent publicly
financed, as the ``cash cows'' of the international publishing houses.

According to estimates, the average total cost (including university housing,
salaries and additional expenses) per published article amounts to
50,000 EUR.  The average profit per published article for
the publishing house, depending on the magazine, is estimated to be
1,000 EUR to 20,000 EUR.  These profits can be maintained only by the
unpaid efforts of authors, reviewers and editors.  Odlyzko
\cite{1997-odlyzko} (see also \cite{1995-odlyzko}), comes to the
following conclusion:

\begin{quote} {\em ``$\ldots$ the monetary cost of the time that
scholars put into the journal business as editors and referees is about
as large as the total revenue that publishers derive from sales of the
journals.  Scholarly journal publishing could not exist in its present
form if scholars were compensated financially for their work.  ''}
\end{quote}

Most publishing houses require the transfer of the exclusive copyright
(not merely the right to use the article) of a scientific report by its
authors and institutions.  This results in the absurd situation that the
very authors and institutions giving away the exclusive copyright to the
journals for free, are less and less able to pay for the rising journal
costs.

The {\it
Association of Research Libraries} speaks openly and candidly of a
communication crisis in the sciences, amounting to the fact that the
scholars have lost control.
These developments have caused the {\it Association of Research Libraries}
to request a radical re-orientation from its customers, the scientist of
North America.
This goes even so far as to suggest very bold moves;
such as the invitation to refuse authorship, as well as a halt on reviews and
editorial activities for the scientific publishing houses.
As a consequence, forms are published, with
which scientists should refuse to review, motivating their denial with
the rising publication costs (see appendix).

However, quite understandably, such attempts show little effect:
each individual scientist would be badly advised to proceed in ``Robin
Hood'' manners  against that very instance which is of crucial
importance in career planning, and which is essential for the official
justification and evaluation of the scientist's research work.
Especially for the novice, not very well established, scientist, this
refusal to publish in peer reviewed journals, would amount to scientific
suicide.  Career decisions are supported by and justified
with publications in as prestigious a journals as
possible.  The higher a marketing value of a journal the better; without
publication in a journal with peer review, there is no career in
science.

\section{E-print servers:  the immature alternatives}

Despite and because of the problems stated above, it should only be a
question of the time until electronic forms of publication will become
generally accepted.  Inevitably, with the  information revolution,  also the forms of scientific
communication will change.
The large publishing houses already
feel the course of the time and react ambivalently:  on the one hand,
they understandably do not like to lose the good business, on the other
hand, an adaptation of the practices appears necessary for future
profits.

Associated with this tactics is a careful, ``snuggling approach'' in
handling ``grassroots'' initiatives such as for instance {\it
arxiv.org}; even if its operators cope with copyright issues rather
vaguely.  In what follows we shall discuss this initiative in some
details; partly because functionally, it is one of the most advanced
ones, partly because it is one with the highest penetration among the
communities involves, and partly because of its apparent success, the
pressing problems appear clearly and visibly.

\subsection{Example {\it arxiv.org}}

{\it arxiv.org} is a reprint and a preprint server, which is freely
accessible publicly to all those who have access to the world wide web.
The speed and simplicity of the information flow, as well as the
relatively small costs of access are important success elements.  {\it
arxiv.org} started as initiative in high-energy physics, and now covers
almost all subfields of physics as well as larger parts of the
mathematical and information sciences.  Configurable daily email
messages are sent out to the subscribers, containing the headline and
abstracts of the articles submitted to the database.  The links in these
emails yield to the manuscript in various full text representations.
This seemingly ideal situation copes with some difficulties, which will
be mentioned below.

It is certainly not the intention of this article to excoriate {\it arxiv.org.} The following
criticism should be understood as a feedback and attempt to make {\it
arxiv.org} even better; maybe also to resolve its functions and content into some
comprehensive archive, which may be able to lead the way to new, more
effective forms of scientific publishing, serving the community even
better that today's archives and peer reviewed journals.

\subsubsection{Copyright skeletons in the closet}

Due to the large popularity of {\it arxiv.org}, and to increase the
dissemination of their research results, many authors do not only submit
their drafts and preprints, but also manuscripts which have already been
published.  Although the layout mostly looks somewhat differently than
``the original article'' in the peer reviewed journal, these copies
contain the identical ``original text;'' as well as all the
illustrations, tables and so on of the original manuscript.  For
authors, this is a reasonable procedure, since self-publication in {\it
arxiv.org} may reach a much larger audience than the journal
publication.
{\it arxiv.org} encourages its authors even to enter the
explicit journal reference.
(Of course, journal references should
be added to a preprint only after final publication.)
Indeed, the quality of the article in {\it arxiv.org} may be even higher
that the one in the paper journal, since errata and further revisions
can be easily included after post-publication.  (All previous version
remain in the database and can be accessed publicly.)

These copyright infringements by many authors make {\it arxiv.org}
vulnerable to lawsuits of publishing houses.  Virtually at any time one
of the large publishing houses, in order to protect its profits and
distribution channels, may sue {\it Cornell University,}
the present operator of {\it arxiv.org,}  because of breach of
copyright.  The American private university {\it Cornell,} in order to
protect itself, might have no other choice that to shut down the
operation entirely. So far, this did not happen; probably for two
reasons:  (i) on the one hand each publishing house cringes because of
the negative publicity in and the affront to the scientific community,
(ii) on the other hand, the losses caused by cancellations of
subscription and substitution by {\it arxiv.org} on the part of the
libraries are still relatively small.  Stated pointedly, at the moment,
the losses in terms of publicity may outweigh the financial gains.
Maybe, in this sense, the publishing houses do not want to wake up the
``sleeping tiger.''  But what would happen when more and more
subscriptions are canceled by the research libraries and profit margins
decrease, appears unforeseeably.

So, unfortunately, without clarifications of legal issues regarding
copyright, the fate of {\it arxiv.org} seems to be uncertain.  Only one
letter of an attorney could cause {\it Cornell} to shut down the {\it
arxiv.org} servers.  This prospect, which particularly would affect the
global physics community, appears as legal-organizational nightmare.
Because {\it arxiv.org} thereby depends on the goodwill of the
publishing houses, whose grace could be lifted at any time according to
the discretion of commercial publishers and their economic
considerations and interests.

\subsection{Obscure procedures}

{\it arxiv.org} still fights with a further problem:  since the authors
self-publish their manuscripts, {\it a priori} it cannot be avoided
that some ``quacks'' self-publish their treatises as well.  It is not
always completely evident who exactly qualifies as ``quack,'' and
whose work does not deserve publication.  Here also {\it arxiv.org} is
hit in full hardness with questions of quality management; one answer
being peer review.

At the moment,
within 24 hours before the final publication, a decision for or against
the final admission into the data base of {\it arxiv.org} is made.  This
is by far quicker that the standard procedures, which may take months to
years.  The decision is made by several, partially anonymous,
moderators, which take over a kind of editorial or publisher role.  {\it arxiv.org}
explicitly states \footnote{ http://arxiv.org/help/general}:
{\em ``We reserve the right to reject any inappropriate submissions.''}

At present, there does not seem to be any kind of official appeals
policy, such as for the journals of the {\it American Physical Society.}
The only possibility remains an informal request by email conversation
with an anonymous censor.  However, the censor's power of decision
remains absolutely.

Rumor has it that ``black lists'' exist, which exclude ``apparent
quacks'' from publication at {\it arxiv.org}.  As a consequence, law
suits have been filed against the operators of {\it arxiv.org} by
authors who were excluded and who insist on their right of free speech
guaranteed by the American constitution also in the scientific domain.

The censorship of the moderators is not completely incomprehensible:
 for legal (``criminal content'') and
technological (``huge data scrap'') reasons, no archive of the world
 might get along without censorship.
It would however be not totally unreasonable to take a
very liberal position in these matters; after all, search engines in
the web make already available, more or less nondiscriminatively,
metainformation and hyperlinks to (parts of) the entire, ``world wild''
uncensored web.

There seems to be no objective demarcation criterion what exactly can be
considered an ``easily recognizable nonsense.''  Due to the informal
character, this term can only be outlined heuristically and concretized
subjectively.  Many old Greeks would have recognized for example someone
as a quack, who would have maintained the ``easily recognizable
nonsense'' (at that that time), that the earth might be a ball,
surrounded by a thin  layer of air,  circling the sun in an almost empty
space; surrounded by hundreds of millions of galaxies, which again consist
of millions of stars.
Almost per definition, revolutionary scientific ideas are difficult to
separate from emanations of ``quacks.''
So, the question arises whether or not  risking
to purge  highly innovative approaches is  worthwhile censoring a few
``quacks.''

Thus the hardly defined standards of self-publication, the alleged black
lists and the arbitrariness of moderators make it necessary to develop
an institutionalization of censorship.

\subsubsection{Quality control}

Closely related to the  obscure censorship procedures just described are
the quality criteria which a publication must meet in order to qualify for {\it
arxiv.org}.  The following declarative statement is at the starting page
of {\it arxiv.org}:  {\em `` The contents of arXiv conform to Cornell
University academic standards.''} The reader is left puzzled what
exactly is meant by ``the Cornell University academic standards;''
nowhere an explanation is given.  Again, the necessity of more
transparent editorial policies is evident.

\subsubsection{Local organization, international goals}

At the moment,
{\it arxiv.org} is operated by Cornell, as it is stated clearly on the
starting page:  {\em ``arXiv is owned, operated and funded by Cornell
University, a private not-for-profit educational institution.  ArXiv is
also partially funded by the National Science Foundation.''} Nobody
alleges {\it Cornell} or any individuals to operate {\it arxiv.org} for
marketing reasons alone, or for any kind of disrespectful intention.  Nevertheless,
the present form of organization of {\it
arxiv.org} as the worldwide archive of literature in the physical and
related sciences, operated by a private American research institution,
appears hardly acceptable.  The archive is just too successful to be
owned and operated by a single institution, which may be much too
susceptible to the possible arbitrariness of
groups  whose self-interests and selfishness might place the own benefit
over the benefit of the international scientific community at large.

\subsection{International e-print server of the UNESCO}

Here we shall briefly outline a proposal for an e-print server operated
by the {\it United Nations Educational, Scientific and Cultural
Organization} (UNESCO), or at least by an international consortium.
Such a service would have to meet the following criteria.

\subsubsection{Copyright issues}

The copyright status of archive entries should be consolidated and
clarified.
A model similar to the GNU
Free documentation License (GDFL) would for instance be conceivable:
 Free availability of the full
contents; any further development of which should be accompanied by a
reference to the original source; a method  which is a standard in the
tradition of established scientific quotation practice as well.

\subsubsection{Clear regulatory schemes}

The complete deletion of articles and authors from the data base (if ever) should
happen according to openly discussed principles.  Here a co-operation
should take place between scientists, layman judges, as well as
philosophers of science, science historians, librarians and specialists
in documentation.

\subsubsection{Interactivity}

Already {\it arxiv.org} has given its authors the possibility to revise
the manuscript, while retaining the older versions.  Readers could be
given the possibility both to anonymous, as well as to not-anonymous
discussion and criticism.  Also, peers could be given the possibility
for ``article sponsorships '', in which other authors signal
their non anonymous agreement with and promotion of an article.

\subsubsection{Evaluation and quality certification}

In the absence of traditional peer review, by far the most
difficult problem is the consequence for career planning and
certification of scientific achievements.  New forms of quantification,
as for instance the evaluation of the access data to a web server, for
example ``hit lists,'' are vulnerable to attack and does not offer any valid criterion.

Mixed scenarios of co-existence between peer
reviewed and non peer reviewed articles together in one big database would also be conceivable.
In this case, certain articles might get certification with special
procedures and certificates, for instance similar to peer review,
which distinguish it over other articles, which do not have this
certification.

One could also conceive of a system of ``peership'' in a somewhat similar
way as in the medieval trade guilds:  if science ``apprentices''
acquire sufficient status, standing and experiences, for instance by
collecting enough certificates, one could declare them to be ``Peer.''
This is not as absurd as it may sound; the old system of academic
lecturership (German ``Dozentur'') works in a similar way.
Once in this status, they could provide certificates, or judge and certify work of
others on the e-print server and elsewhere.

Alas, if one perceives scientific career planning in another light,
these problems appear not so difficult and all-important:  often,
certifications and quantitative criteria are used merely
for the post-justification
and ``objectification'' of subjective opinions and career decisions.

\section{Does peer review more good than bad?}

In the long run, this question probably cannot be answered with a
clear ``yes'' or  ``no.'' The connections with scientific career
planning and business are too complex.  Everyone should form an own
judgment.

The author recognizes the great advantages and assistance, which peer
review offered to him throughout his scientific activities;
yet he believes that these advantages were at least partly
nullified by the often senseless delay of publications, and sometimes
associated with a distortion of contents and vain expenditures.  This
may be particularly the case for ``original'' and innovative contents,
and may not be so urgent for well established research topics, where the
quality improvement gained by peer review may be marginal anyhow.

Maybe the most decisive factor will be money;
the libraries and public households, as well as the autonomous
universities will not want or simply will not be able to pay the
exploding costs of the peer reviewed printed media.  It is unlikely that
the scientific community will bring down peer review because of a
widespread refusal to review and edit; rather there will be a constant,
concealed deterioration of the quality of unpaid and unrewarded reviews.

Eminent scientific European organizations, such as for instance the {\it
Centre National de la Recherche Scientifique (CNRS)}, the {\it
Max-Planck-Gesellschaft (MPI)}, or the {\it Deutschen
Forschungsgemeinschaft (DFG)}, just to enumerate a few, begin to
acknowledge the new developments, which are tied to the new electronic
media, and which are revolutionizing scientific publishing as these lines
are written.  In their ``Berlin Declaration on Open Access to Knowledge
in the Sciences and Humanities,'' they express their commitment to open
archives, as well as their determination to honor such publications
for career planning and quality management
\footnote{
http://www.mpg.de/pdf/openaccess/BerlinDeclaration\_en.pdf }:
\begin{quote} {\em ``[[$\ldots$]] Obviously, these developments will be
able to significantly modify the nature of scientific publishing as well
as the existing system of quality assurance.

[[$\ldots$]] Therefore, we
intend to make progress by [[$\ldots$]] encouraging our
researchers/grant recipients to publish their work according to the
principles of the open access paradigm  [[$\ldots$]] developing means
and ways to evaluate open access contributions and online journals in
order to maintain the standards of quality assurance and good scientific
practice   [[$\ldots$]] advocating that open access publication be
recognized in promotion and tenure evaluation.  $\ldots$'' } \end{quote}

It remains to be seen, if and how fast these organizations will actually adopt the principles
to which they have committed themselves.
Perhaps
courageous prominent and financially secured authors with high
status should make a first step.  This would make necessary a type of
open archives which makes sure that no ``backup'' from conventional
publishing is necessary.
However, at the moment, despite the early success of
{\it arxiv.org} and others, no such commitment exist.

As concerns the funding of research proposals, radically different
allocation and distribution strategies of are conceivable and appear not
totally unreasonable; one possibility would be to distribute about 70 \%
of the funds via traditional peer review channels; 20 \% over a system
of lay judge; as well as 10 \% completely at random.
Such a strategy
would have to be accompanied and adapted by additional evaluations,
which would have to take place again via scientists as well as
layman judges.

Finally again an aspect of scientific should should be mentioned, which one
cannot value high enough:  Science can remain only alive and
productive if it is operated with passion and joy and is fun to
pursue.  It makes little sense to obstruct young people with
little pay, who are willing
to dedicated much of their lifetime to the scientific progress,
through malign treadmills organized around the presently executed ``objective''
schemes of peer evaluations.  That will simply not work, and will result
in a waste of taxpayer's money.

It is really amazing, how indifferent the scientific establishment at large,
reacted to the findings which clearly indicated substantial problems
for instance in the funding of research projects.
This neglect, for instance of the Cole report
\cite{1981-cole},
suggests that the stakes of particular interest groups might be at work;
stakes which might reflect self-dedication and self-interests, but might
not serve the best purposes of the societies or the trusts which raise the money.

In the long run, everything is subjected to a historical change; this
applies also to scientific publications and the methods how scientific
achievements are evaluated.  One is reminded of president Roosevelt's
Address to the U.S.  Congress in 1941 \cite{RooseveltFD}:  {\em ``we
have been engaged in change -- in a perpetual peaceful revolution - a
revolution which goes on steadily, quietly adjusting itself to changing
conditions.''}

\appendix

\section{Anecdotes}

The following authentic anecdotes were anonymized.  This is neither a
comprehensive collection, nor is it representative.
These anecdotes should be understood as just a few of the more or
less funny stories almost any author, reviewer or editor can tell.

\subsection{} A publisher received two contradicting peer reviews of an
article:  The first report found that the idea was pointless and
unrealistic, however its formal elaboration correct.  The second referee
found that the idea was extremely interesting, yet its formalization was
bad.

\subsection{} In his review, a peer stated that he would not recommend
 publication if the article came from a less established author and
recommended the article with these {\it provisos}, for reasons of seniority.
Afterwards, the ``anonymous'' peer reviewer contacted the author
privately and attempted to direct him to his own work.

\subsection{} After his retirement, a very renown author tried to
publish some of his research under other ``brand'' names.  He wanted to test
the system.  Thereby, he failed completely; most of the works of the
seemingly ``unknown'' authors got rejected immediately.  They were
accepted with a breeze under the author's ``brand'' name.

\subsection{} Another, very renown, author expressed his unwillingness
to expose himself to peer review and stressed that he does not publish
any more in peer reviewed journals.  He had the feeling that the referee
reports mostly missed the point and were mean and not helpful at all.  He had
enough invitations to write contributions for conferences and
anthologies, for which the editors usually stated their criticism much
more carefully.  (See also the Einstein anecdote.)

\subsection{} A team of researchers decided to publish a very important
and original result not in a peer reviewed journal because of the
danger of the delay and refusal by peer review, but rather ``hid it'' in
a conference volume.  This article was then cited by and based upon a
vast number articles in peer reviewed high-ranking journals.

\subsection{} Scientist A called Scientist B and asked for assistance.
B should write an article, in which A's point of argument was defended
against criticism of author C. B agreed, and wrote the article after
consulting A many times.  The first round of reviews of B's article
claimed that B understood nothing at all about A's intentions
(note:  B was animated and consulted by A).  Therefore, the
article got rejected immediately.  The second round of Referees produced
a nasty referee report, in which the reviewer again did not deal with
details, but called the paper ``perverse.''  On the basis of this judgment,
the paper was finally rejected. It was accepted almost immediately by another journal.

\subsection{} The publisher of one `` Letter'' journal, which, according to
its own understanding, is dedicated to the ``rapid dissemination'' of
scientific results, needed one and a half months just to decide that the
length of the article exceeded the permitted length by five per cent.
The article was rejected for that reason without further review.

\subsection{} On the door of a colleague a poster declares:  {\em
``Save a Tree, Reject a Paper.''}

\section{Sample letter against reviewing activity}

 The {\it Association of Research Libraries,} in the framework of her {\it Create}-initiative, has issued the following sample letter
\footnote{
http://www.arl.org/Create/faculty/tools/letters/letters1.html
und \\
http://www.arl.org/Create/faculty/tools/letters/letters2.html
}:

\begin{quote}
{\em
Dear -------:

It is with great regret that I notify you that I am no longer able to serve as a reader/referee for articles submitted to Title of Journal.

I am brought to this decision because your pricing policy for this journal is at odds with a fundamental value of scholarship, to make scholarly research as widely available as possible. Because of the journal's extraordinarily high cost and astonishingly high annual price increases, it has effectively been placed out of reach of many of my colleagues whose libraries can no longer afford it.

      I feel that you have lost touch with the core purposes of scholarly communication, and I cannot, in conscience, participate in an enterprise that apparently values profit more than the goals of scholarship.

      Moreover, I shall now seek to support, through my submissions and my reviewing activities, alternatives to Title of Journal that maintain affordable costs, as well as cost increases that are clearly related to actual production costs and added value -- in short, costs that promote the widest possible availability of my work and the work of my colleagues.

      Should you change your pricing policies so that they are more in line with scholarly values, please let me know.

      Sincerely,
}
\end{quote}

%\bibliography{svozil}
%\bibliographystyle{apsrev}

\end{document}